\documentclass[%
 aip,
 jmp,%
 amsmath,amssymb,
preprint,%
]{revtex4-1}

\usepackage{graphicx}
\usepackage{dcolumn}
\usepackage{bm}
\usepackage{color}

\graphicspath{ {figures/} }

\begin{document}

\preprint{AIP/123-QED}

\title{Voltage Driven, Local, and Efficient Excitation of Nitrogen-Vacancy centers in Diamond}

\author{D. Labanowski}
 \thanks{These two authors contributed equally}
 \affiliation{Department of Electrical Engineering and Computer Sciences, University of California, Berkeley}
 \affiliation{Intelligence Community Postdoctoral Research Fellowship Program, University of California, Berkeley}
\author{V. P. Bhallamudi}
 \thanks{These two authors contributed equally}
 \affiliation{Department of Physics, The Ohio State University}
\author{Q. Guo}
 \affiliation{Department of Physics, The Ohio State University}
 \author{C. M. Purser}
 \affiliation{Department of Physics, The Ohio State University}
 \author{B. A. McCullian}
 \affiliation{Department of Physics, The Ohio State University}
\author{P. C. Hammel}
 \affiliation{Department of Physics, The Ohio State University}
\author{S. Salahuddin$^*$}
 \email{sayeef@berkeley.edu}
 \affiliation{Department of Electrical Engineering and Computer Sciences, University of California, Berkeley}

\date{\today}

\begin{abstract}
Magnetic sensing technology has found widespread application in industries as diverse as transportation,\cite{cheung2005traffic, treutler2001magnetic} medicine,\cite{hamalainen1993magnetoencephalography, hamada1999prenatal} and resource exploration.\cite{robinson2008advancing, goetz1983remote, nabighian2005historical} Such use cases often require highly sensitive instruments to measure the extremely small magnetic fields involved, relying on difficult to integrate Superconducting Quantum Interference Device (SQUID)\cite{weinstock2012squid} and Spin-Exchange Relaxation Free (SERF)\cite{kominis2003subfemtotesla} magnetometers. A potential alternative, nitrogen vacancy (NV) centers in diamond, has shown great potential as a high sensitivity and high resolution magnetic sensor capable of operating in an unshielded, room-temperature environment.\cite{rondin2014magnetometry} Transitioning NV center based sensors into practical devices, however, is impeded by the need for high power RF excitation to manipulate them.\cite{teale2015magnetometry, clevenson2015broadband, brenneis2015ultrafast} Here we report an advance that combines two different physical phenomena to enable a highly efficient excitation of the NV centers: magnetoelastic drive of ferromagnetic resonance (FMR) and NV-magnon coupling. Our work demonstrates a new pathway to combine acoustics and magnonics that enables highly energy efficient and local excitation of NV centers without the need for any external RF excitation, and thus could lead to completely integrated, on-chip, atomic sensors.
\end{abstract}

\maketitle

\label{(sec:introduction)}
Experiments investigating high-sensitivity magnetometry with NV centers commonly require the application of a microwave frequency excitation on the order of 1-10 Watts.\cite{teale2015magnetometry, clevenson2015broadband, brenneis2015ultrafast} The need for such large excitation power not only complicates the integration of the sensors in small footprint, but also has the potential to substantially perturb the environment that it is trying to measure. For example, a 10 Watt microwave excitation traveling in a standard 50 $\Omega$ stripline creates a magnetic field of approximately 90 $\mu$T at a distance of 1 mm, thus making it impossible to take advantage of the intrinsic ability of NV centers to detect fields of the order of 10's of fT in an unshielded environment.\cite{wolf2015subpicotesla}

One potential mechanism for localizing the influence of the incident RF power to the diamond NV centers is by leveraging the recently observed interaction between NV centers in diamond and a proximal resonating ferromagnet.\cite{wolfe2014off, wolfe2016spatially, page2016optically, du2017control, van2015nanometre} The time-varying excitations responsible for the NV-FMR coupling are spatially periodic magnons, whose magnetic effects should be entirely confined to within a few wavelengths (on the order of micrometers or less for these systems). Commonly, however, such systems run into the same problems of high-power RF excitation - with studies utilizing more than 15 Watts of microwave power in order to obtain low-noise measurements.\cite{du2017control, van2015nanometre}

\begin{figure}
	\centering
	\includegraphics[width=1\textwidth]{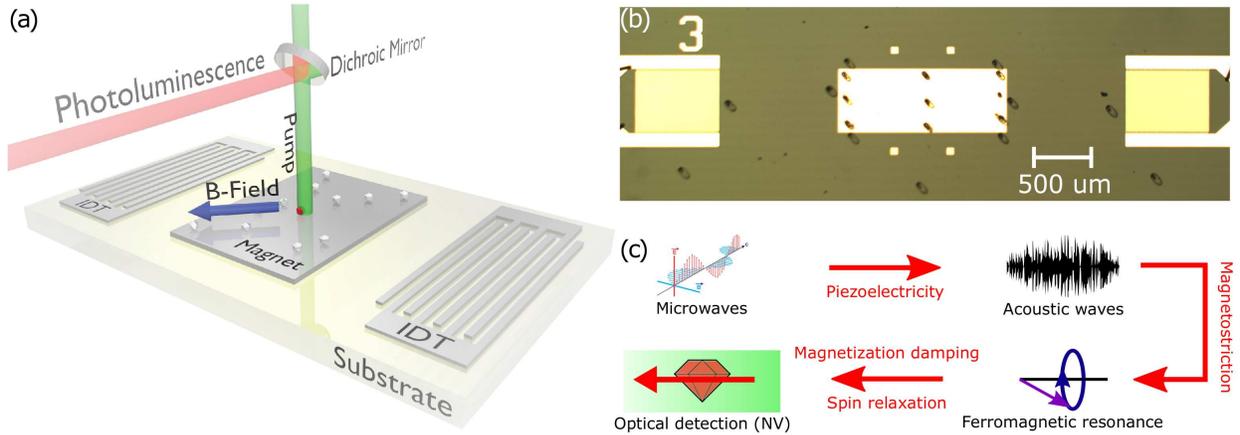}
	\caption{Measured NV-ADFMR system: (a) Schematic diagram of experimental sample and optical excitation / detection scheme. The magnetic field was applied at 45$^\circ$ in-plane from the SAW propagation direction for all optically-detected measurements. Small particles on the magnetic pad indicate deposited nanodiamonds. (b) Photograph of measured device shows interdigitated transducers and the magnetoelastic film. Dark spots on film and substrate are clusters of nanodiamonds. (c) Diagram of energy flow in the system, showing transduction methods between the different components of the sample. Microwave electrical energy is converted into acoustic energy via piezoelectricity, which then drives magnetic precession. As this precession damps, it generates magnons that couple to the NV centers, modulating their photoluminescence.}
	\label{fig:schematic}
\end{figure}

Here we report an advance that combines two different physical phenomena to enable a highly efficient excitation of the NV centers (see Figure \ref{fig:schematic}). The energy flows in this combined system are outlined in Figure \ref{fig:schematic}(c). The first of these is the recent demonstration that FMR in a thin ferromagnetic film can be excited using magnetoelastic interaction with a piezoelectric material. \cite{PhysRevLett.106.117601, gowtham2015traveling, labanowski2016power, labanowski2017effect}  This allows for the excitation of a purely voltage driven FMR. The magnetoelastic interaction transduces acoustic excitations into an {\it internal effective} magnetic field within the ferromagnet. As a result, the field is extremely local and resides within the ferromagnetic film atop the piezoelectric material. In addition, this effective magnetic field is several orders of magnitude more efficient than traditional FMR excitation via Oerstead fields in a stripline,\cite{labanowski2017effect} reducing the input power requirement by the same amount. Once the ferromagnet is excited into FMR, the highly local interaction between the NV centers and spatially periodic magnons can now be employed to excite the NV centers. Therefore, the entire interaction from piezoelectric to NV centers is highly local. This, combined with the fact that the input power requirements are substantially reduced, leads to minimal perturbation of the surrounding environment.

\label{(sec:experiment)}

A schematic diagram of our experiment can be found in Figure \ref{fig:schematic}(a), and an optical image of a measured device can be found in Figure \ref{fig:schematic}(b). The system consists of an acoustically driven ferromagnetic resonance (ADFMR) device where microfabricated resonant interdigitated transducers (IDTs) launch surface acoustic waves (SAWs) in the piezoelectric LiNbO$_3$ substrate when driven with a microwave voltage. A ferromagnetic pad, either cobalt or nickel (with a thickness of 20 nm), is located on top of the substrate. More details on the fabrication of these devices can be found in existing literature\cite{labanowski2016power, labanowski2017effect}. Nanodiamonds containing NV centers were deposited at multiple locations on top of the ferromagnetic pad to enable localized optical detection of the acoustically driven FMR. These spots can be seen as the dark circles in the optical image.

\begin{figure*}
	\centering
	\includegraphics[width=0.95\textwidth]{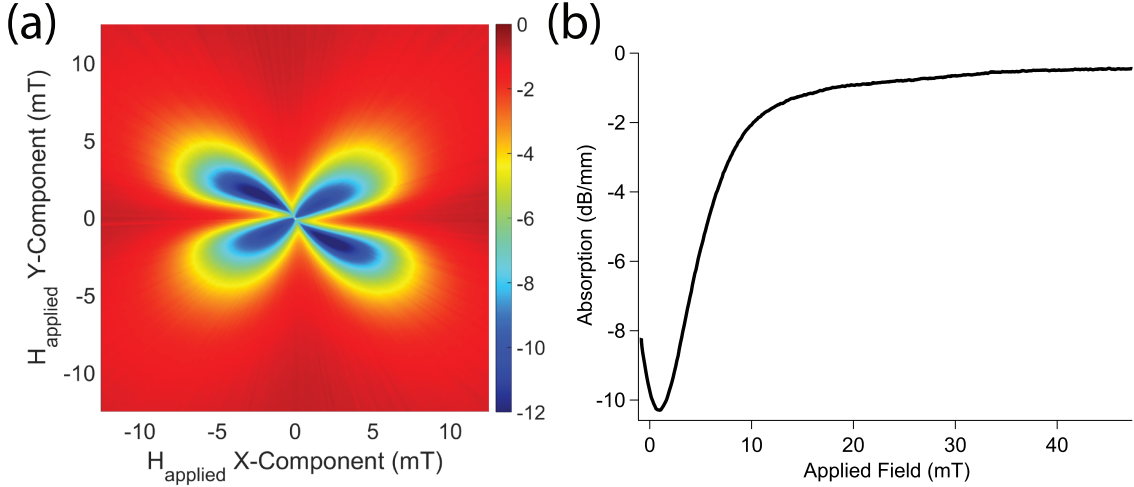}
	\caption{Power absorption in acoustically driven ferromagnetic resonance: (a) Plot of power absorption as a function of applied magnetic field for a 20 nm nickel ADFMR device at 1429 MHz. The x-component of the field is taken to be parallel to the direction of SAW propagation, and the y-component is in-plane and perpendicular to the direction of SAW propagation. The color bar indicates absorption in dB/mm. (b) Line cut along 45$^\circ$ showing a large field-dependent absorption in the orientation used for NV coupling measurements.}
	\label{fig:ADFMR}
\end{figure*}

Figure \ref{fig:ADFMR} shows ADFMR absorption data collected for a 20 nm nickel device similar to those used in this study at 1429 MHz. The dependence of absorption on applied magnetic field and field angle can be seen in Figure \ref{fig:ADFMR}(a). This data was taken using the same procedures described in previous work by the authors.\cite{labanowski2017effect} Notably, a time-gating technique was employed to more accurately measure the attenuation of the surface acoustic waves and to reject signals caused by spurious microwave radiation from the IDTs. Figure \ref{fig:ADFMR}(b) shows a line cut of the field dependence 45$^\circ$ from the SAW propagation direction. The microwave absorption can be observed to be extremely large for such a low-frequency excitation - several orders of magnitude more than a traditional stripline FMR excitation.\cite{maksymov2015broadband}

The ADFMR phenomenon occurs due to a combination of piezoelectric and magnetoelastic interactions. An RF voltage applied to an IDT first generates surface acoustic waves. These waves generate a time-varying strain that can alter the magnetocrystalline anisotropy of a magnetoelastic ferromagnet via the Villari effect and generate an effective magnetic field internal to the magnet\cite{PhysRevLett.106.117601}. This effective field is capable of driving the system into FMR. The effect has been quantified by measuring the attenuation of the traveling acoustic wave. A more detailed explanation of this phenomenon can be found in existing literature\cite{labanowski2016power, PhysRevLett.108.176601, gowtham2015spin, PhysRevB.86.134415, labanowski2017effect}.

As the precession generated by ADFMR is damped, it emits spin waves over a range of frequencies\cite{page2016optically, du2017control}. The NV center is a spin 1 defect whose photoluminescence (PL) is dependent on its spin-state and is hyperpolarized into its 0-state by the 532 nm pump laser\cite{harrison2006measurement}. The incoherent spin waves emitted by the precession as it damps can be resonant with the NV, causing the NV spins to relax. NV PL intensity is spin-dependent, being high for  0-state and low for $\pm 1$ state. Thus, as the NV spins relaxes due to FMR, their PL intensity changes and can be recorded.

\begin{figure}[tb]
	\centering
	\includegraphics[width=0.9\textwidth]{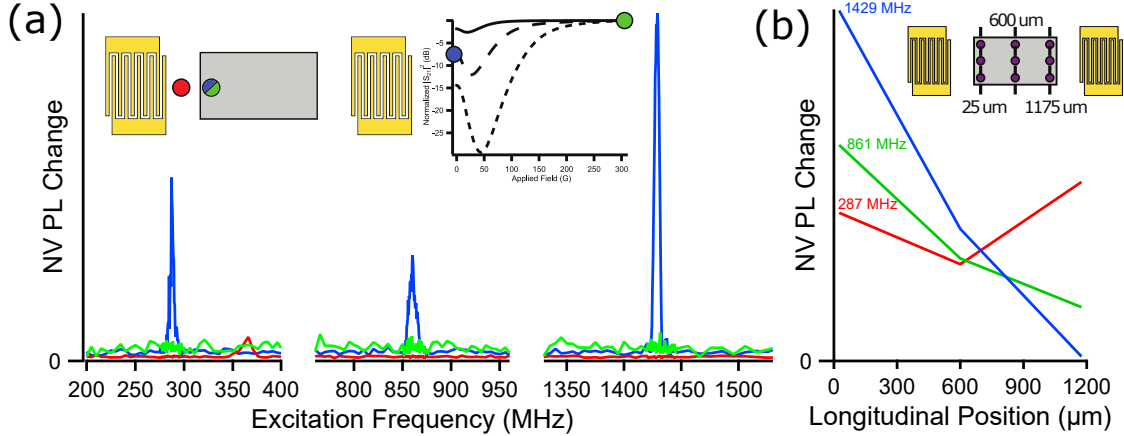}
	\caption{Frequency and spatial dependence of ADFMR-NV coupling in nickel devices: (a)  Change in NV center PL normalized to the DC level for nanodiamonds both on and off the ferromagnetic pad. NV centers located off the ferromagnetic pad (red) and NV centers on the pad with a high (35.8 mT) applied bias field (green) show no change in photoluminescence. Only NV centers on the pad with 0 applied bias field (blue) show a notable photoluminescence change. All measured NV photoluminescence change outside of the shown frequency range is within noise. Left inset - position of the NV centers on the device. Right inset - power absorption of the ferromagnetic film along the applied field direction at the field values of interest. The colors in the insets correspond to colors in the figure. The peaks in the NV center signal align with the first, third and fifth harmonics of the IDTs. (b) NV PL change in a 20 nm nickel sample as a function of longitudinal position from the edge of the ferromagnet closest to the excitation IDT. Measurements were performed with the drive frequency set to the first (red), third (green), and fifth (blue) harmonics of the IDTs and at zero applied magnetic field. Inset - schematic illustration of nanodiamond positions.}
	\label{fig:couplingDep}
\end{figure}

Figure \ref{fig:couplingDep}(a) shows the change in NV center PL under different conditions for a nickel device. Shown are frequency sweeps done around the first, third and fifth harmonics (287, 861, and 1429 MHz) of the IDTs; the IDTs only allow odd harmonics to be transmitted as they are patterned with a 50\% metalization ratio\cite{9780444888457}. The scans shown in Figure \ref{fig:couplingDep}(a) are representative scans to demonstrate the optical detection of ADFMR. The amplitude of the peaks can vary between samples and between nanodiamond clumps on the same sample, in addition to the variation due to position and frequency. Near the edge of the ferromagnetic pad closest to excitation IDT, at higher frequencies, the average fractional change is approximately 1\%. All data for PL change presented in this paper is fractional change, i.e., lock-in voltage/DC voltage.

Clear peaks in the optical signal can be seen for nanodiamonds at zero magnetic field and on the Ni pad (blue). The NV spins by themselves should not have any signal at 287 and 861 MHz since they are far away from any NV resonances in zero field. However, at 1429 MHz a direct-drive signal can be observed since this coincides closely with the NV's excited state resonance frequency. To check the origin of the signal we performed a control measurement away from the ferromagnetic pad (red), which shows no signal even at 1429 MHz, even though this spot was closer to the input IDT and should have a larger amplitude of acoustic waves (and spurious microwaves). Thus we can rule out any spurious interaction of the acoustic waves or microwaves directly with the NV in our experiment. While strain waves can couple to the NV center\cite{macquarrie2013mechanical, golter2016optomechanical}, we do not see any evidence for this interaction in our experiment, most likely due to the very weak mechanical coupling between the substrate and the nanodiamonds. To further verify the FMR origin of the peaks scene in the blue data, we measured the signal at higher field (green). A field strength of 35.8 mT is sufficient to bias the magnetic films out of resonance. As it can be seen, we observe no PL signal at these fields. Thus our control measurements verify that the peaks scene in the blue data are due to FMR.

We show the local nature of the drive by recording spatially localized data across the ADFMR device. Figure \ref{fig:couplingDep}(b) shows the change in fractional NV PL at zero field for the 20 nm nickel device as a function of longitudinal position, i.e., distance from the edge of the nickel pad closest to the excitation IDT along the SAW propagation direction. The data presented here is an average of three spots that are the same distance from the edge of the nickel pad (see inset). This averaging helps mitigate the effect of inhomogeneity in nanodiamond spots at various locations. The distance between the NV centers and the ferromagnet determines coupling efficiency, and thus inhomogeneity arising from the deposition process can lead to a noisy signal. These direct spatial measurements performed via coupling to NV centers corroborate previous ADFMR studies that show similar qualitative behavior as a function of position\cite{labanowski2016power} and frequency\cite{labanowski2016power, PhysRevLett.108.176601, gowtham2015spin, PhysRevB.86.134415}.

\begin{figure*}[tb]
	\centering
	\includegraphics[width=1\textwidth]{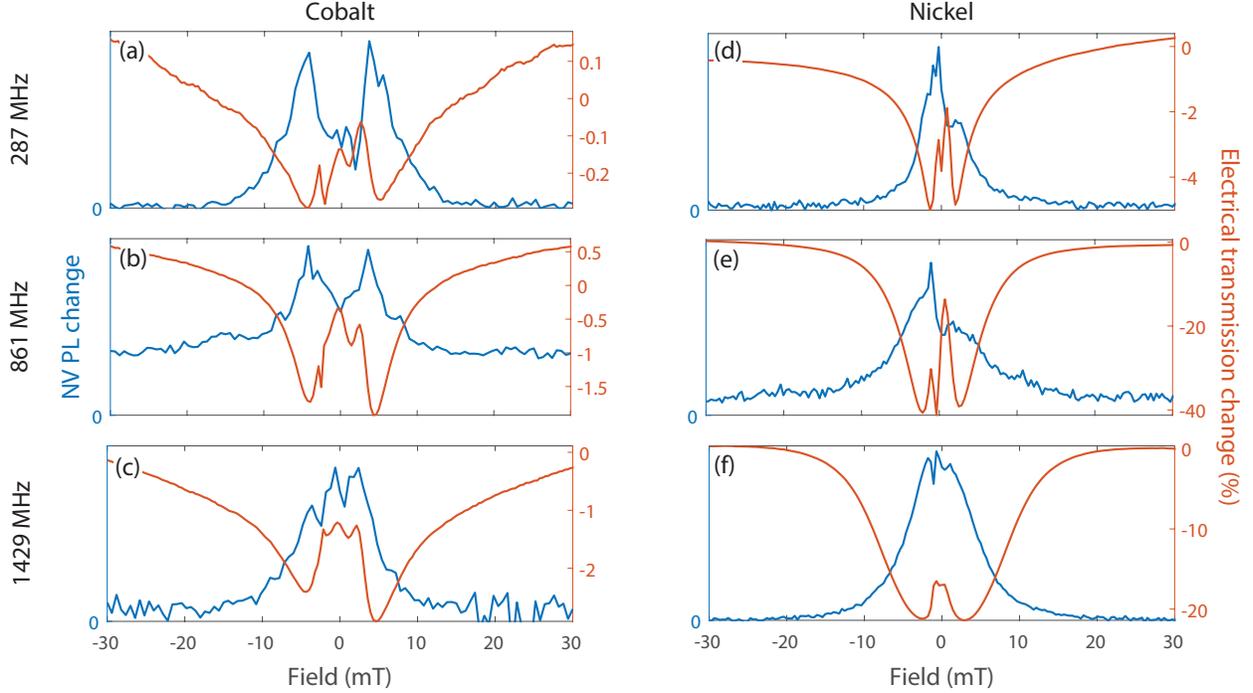}
	\caption{Magnetic field dependence of ADFMR-NV coupling: Microwave power absorption and NV PL change as a function of magnetic field in nickel and cobalt samples. The drive frequency for the ADFMR devices was set to the first, third, and fifth harmonics of the IDTs for these measurements. A clear correlation between power absorption and NV PL change can be observed.}
	\label{fig:fieldDep}
\end{figure*}

We now provide further evidence for optical detection of ADFMR by performing field sweeps around these IDT harmonic frequencies for both cobalt and nickel devices. These data are presented Figure \ref{fig:fieldDep}, where we show change in the detected electrical signal in addition to the optical signal. Correlation between the optical signal and the transmission signal can be seen, especially in the case of cobalt. No peaks in the optical signal were seen in nanodiamonds not located on the ferromagnetic pads. It should be noted that while we have previously detected FMR in cobalt using our relaxation-based technique, this is the first report of optical detection of FMR in nickel using NV centers. The peak heights of the optical signal are comparable in both cobalt and nickel, even though the absorption signal is an order of magnitude (or more) larger in nickel films. This would imply that the coupling between nickel and NV centers is much smaller than for the case of cobalt. The reason for dependence of the coupling on the magnetic material is currently unknown.

\label{(sec:conclusion)}

To summarize, we demonstrate a new pathway for off-resonant excitation of NV centers in diamond. A voltage driven acoustic wave can generate FMR in a ferromagnetic thin film by exploiting magnetostriction. This FMR in turn couples with NV centers through spatially varying magnons. Both coupling mechanisms are highly local. In addition, the magnetostriction driven FMR is a few orders of magnitude more efficient as compared to conventional Oersted field driven FMR.\cite{labanowski2017effect} Given optimized IDT designs that do not suffer from large insertion losses ($\approx$ 35 dB when operating at 1429 MHz for the devices used in this study), it should be possible to drive such a coupled system at extremely low power levels. Thus, the combination of acoustics-to-magnonics-to-NV centers allows for an extremely local and energy efficient way to excite the NV centers without perturbing the surrounding environment. Although in this first demonstration we have used PL to probe the NV centers, it is conceivable that with proper optimization it should be possible to detect them purely electrically by observing the power absorption itself. The choice of atomic levels are also not limited to NV centers - any other defect center with zero field splitting should show similar behavior. Further work should focus on these possibilities as well as the fundamentals of the coupling mechanism between the ferromagnet and the NV centers. Advances in these areas should lead to completely integrated, electrically driven and interrogated, atomic sensors that work close to their intrinsic limits.

\section*{Methods}
\label{(sec:methods)}
ADFMR devices were fabricated on commercially-available Y-Cut LiNbO$_3$ substrates (MTI Corp). Patterning was performed using standard g-line photolithography. Al contacts were deposited using thermal evaporation, and the nickel and cobalt films were deposited using e-beam evaporation and sputtering, respectively. Commercial nanodiamonds containing NV centers (Ad\'amas Nanotechnologies) were deposited using laser-pulled glass pipettes at multiple locations on top of the ferromagnetic pad. A signal generator was used to generate the microwave excitations, and electrical transmission measurements were performed both using both a time-gating technique in conjunction with a spectrum analyzer\cite{labanowski2016power, labanowski2017effect} as well as by using a lock-in amplifier and an RF power diode.

Optical PL measurements were performed using a home-built NV characterization setup.  A 532 nm laser was used to excite the NV centers and the resulting PL was measured using a photodiode. A lock-in measurement was performed by modulating the amplitude of the acoustic waves in order to enable the measurement of small changes in the NV PL. Details of the measurement of FMR using NV centers can be found in existing literature\cite{wolfe2014off}. All optical measurements in this work were performed with the external bias field oriented at 45$^\circ$ in-plane from the SAW propagation direction.

\section*{Acknowledgements}
\label{(sec:acknowledgements)}

This research was supported by an appointment to the Intelligence Community Postdoctoral Research Fellowship Program at the University of California, Berkeley, administered by Oak Ridge Institute for Science and Education through an interagency agreement between the U.S. Department of Energy and the Office of the Director of National Intelligence. Funding for the research at Ohio State University was provided by the Army Research Office through Grant W911NF-16-1-0547. The authors would like to thank Dr. Charles Henry Lambert for assistance with sample preparation and Dr. Praveen Gowtham and Niklas Roschewsky for useful discussions.

\section*{Author Contributions}
\label{(sec:contributions)}

D.L., V.P.B., Q.G., C.M.P., and B.A.M. performed measurements and analyzed data. D.L. worked on device fabrication. V.P.B., B.A.M., Q.G., and C.M.P. built the measurement setup. P.C.H. and S.S. initiated the research. D.L. and V.P.B. wrote the manuscript. All authors read and commented on the manuscript.

\bibliography{references}
\end{document}